# Combining Physics and Mathematics Learning: A Taylor Series Analysis of an Oscillating Magnetic Field

S. Ortuño-Molina[a], A. Garmendía-Martínez[a], P. Fernández de Córdoba[b], J. C. Castro-Palacio[a], J. A. Monsoriu[a], and F. M. Muñoz-Pérez[a,c]

[a]Centro de Tecnologías Físicas, Universitat Politècnica de València, Camino de Vera s/n, València, 46022, Spain
[b]Instituto Universitario de Matemática Pura y Aplicada, Universitat Politècnica de València, Camino de Vera, s/n, València, 46022, Spain
[c]Institut Universitari de Ciencia dels Materials (ICMUV), Universitat de Valencia, 46100, Burjassot, Spain.

## Abstract

In this work, we present a simple and low-cost experiment designed to study the oscillations of the magnetic field created by a cylindrical magnet under two different conditions: far and short distances from the magnetic sensor. A Taylor's series expansion of the magnetic field function has been done to study the convergence of the polynomial series to the real field in both situations. To carry out the experiment, a small cylindrical magnet has been attached to an oscillating and well-known spring-mass system. The resulting oscillating magnetic field has been registered with the smartphone by using the magnetometer sensor. A very good agreement has been obtained between the theoretical model for the magnetic field and the experimental data collected with the sensor located near and far from a cylindrical magnet and along its longitudinal axis.

# 1. Introduction

Current smartphones are equipped with a wide range of sensors, including accelerometers, gyroscopes, barometers, magnetometers, and light intensity sensors, among others. These devices are not only useful for communication and work but also have applications in recreational activities. However, their potential in the educational field, specifically in the experimental area of General Physics, has been little explored. The possibility of reducing laboratory costs, their portability, allowing experiments to be conducted in disadvantaged environments, and, above all, the increased motivation of students toward practical work due to their advanced measurement and data recording capabilities make these sensors valuable tools that deserve deeper analysis.

The use of these sensors in smartphones is not limited to easy access and good measurement precision. They also enable students to interpret physical phenomena more clearly and immediately (Hochberg et al., 2018). There are free applications, such as Physics Toolbox Suite (Vieyra, 2020) and Phyphox (Staacks et al., 2018), available for both Android and iOS, that allow real-time data visualization or export for later analysis in spreadsheets. This accessibility not only optimizes the efficiency of experiments but also strengthens the interaction between students and the physical phenomena being investigated, making the learning process more dynamic and understandable.

Among the most useful sensors for educational purposes is the magnetometer or magnetic field sensor. It has been employed in previous work to measure magnetic fields generated by small spherical and ring magnets, as well as to evaluate magnetic permeability (Wannous et al., 2023), the magnetic field of a circular loop (Ogawara et al., 2017), and the field induced by a coil (Monteiro et al., 2017). Additionally, it has been widely used to study the mechanical aspects of moving objects (Dumrongkitpakorn et al., 2023). Although there are already studies on the mathematical modeling of magnetic fields (Erol and Kara, 2023), these present significant differences from the approach we propose in this work. So far, there is no significant work on how many terms of Taylor's series are necessary to accurately describe the transition between two very different conditions, such as the near and far fields created by an oscillating magnet. The concept of oscillation related to the use of the magnetometer has also been addressed in previous research (Westermann et al., 2022; Listiaji et al., 2023), reinforcing the relevance of this sensor in the educational field.

The use of smartphone sensors in physics teaching extends beyond the magnetometer. For example, the acceleration sensor or accelerometer has been employed to study phenomena such as free fall and the motion of a simple pendulum (Vogt & Kuhn, 2022 a, b). Additionally, it has been used to analyze free harmonic and damped motion in springs, vibration modes in coupled oscillators (Castro-Palacio et al., 2013 a, b), and mechanical beating (Gimenez et al., 2017).

Another widely used sensor is the gyroscope, which, when combined with the accelerometer, has facilitated studies of angular momentum conservation (Shakur, 2013), rotational energy in a physical pendulum, and the relationship between angular velocity and centripetal acceleration (Monteiro, 2014).

The microphone and loudspeaker are also popular tools, either separately or in combination. They have been used to study sound waves, including determining the speed of sound in a gas (Parolin and Pezzi, 2013); analyzing acoustic waves and beats (Kuhn & Vogt, 2013 a, b), and studying phenomena such as the Doppler effect (Gómez-Tejedor et al., 2014), sound interference (Polak, 2016), the inverse-square law of sound intensity (Marín-Sepúlveda et al., 2014), or the the study of the frequency response and resonance in a series RLC circuit (Torriente-García et al., 2023), amongst others.

Additional sensors, such as the ambient light sensor (Sans et al., 2013; Salinas et al., 2018a) and the barometer (Salinas et al., 2018b), have also proven to be robust and effective in a variety of experiments, making them ideal for classroom physics lessons.

A comprehensive resource showcasing the use of various smartphone sensors for innovative physics teaching is the book by Jochen Kuhn and Patrick Vogt (2022), which features more than 70 experiments utilizing smartphones as mini-laboratories.

A common challenge in teaching Physics and Mathematics is that students tend to learn both disciplines in isolation, without recognizing the deep connection between them. This article seeks to demonstrate interdependence through the study of an oscillating magnetic field generated by a small magnet along its longitudinal axis under two different conditions. Although students are taught that the Taylor expansion around x=0 of a mathematical function can be expressed as an infinite sum of polynomial terms (Riley et al., 2006), its practical application in modeling engineering systems is rarely addressed and limited to the first order of the Taylor series (Bissell, 2023). A key question is: how many terms are necessary for the polynomial to approximate the original function with a given level of accuracy?

In this work, we propose a low-cost, portable, and relevant experiment for students in the final years of Secondary Education and the early years of Engineering degrees. The experiment focuses on modeling an oscillating magnetic field through Taylor series expansion under two conditions: when the magnetic field is measured near the source (near field) and when it is measured at a greater distance (far field). In the case of the far field, where the amplitude of the oscillating system is significantly smaller than the equilibrium distance between the magnet and the sensor, the disturbance generated in the system behaves sinusoidally, so only two terms of the series expansion are required to adequately model the behavior of the oscillator. However, in the case of the near field, where the equilibrium distance and the oscillation amplitude are comparable, although the behavior of the system remains continuous, bounded, and periodic, it loses its characteristic sinusoidal shape. This means that more terms of the expansion are needed to accurately model the behavior of the oscillator, as will be detailed in the results section. This approach not only introduces advanced mathematical concepts but also allows for exploration of the modeling of physical systems, aspects that are rarely addressed in introductory physics courses.

## 2. Theoretical model for the magnetic field of a cylindrical magnet

The axial component of the magnetic field generated by a cylindrical permanent magnet of radius $R$ and height $L$ is given in the literature by the following approximate expression for distances $z \gg L$:

$$B(z) = \frac{\mu_0 m}{2\pi z^n} \qquad (1)$$

where $\mu_0 = 4\pi \cdot 10^{-7}$ N·A$^{-2}$ is the magnetic permeability of the vacuum, $m$ the modulus of the magnetic dipole moment $m = M \cdot V$, and $M$ is the magnetization of the magnet with volume $V = \pi \cdot R^2 \cdot L$. The parameter $z$ is the axial distance measured from the mid-point of the magnet to the magnetometer.

In this first part, and to make sure of the dependence of the field with the inverse of the distance cubed, we proceeded to the experimental determination of this exponent $n$ using the magnetometer of the cell phone through a very simple method. For this purpose, the free Phyphox app and an Apple iPhone xR were used. We located the position of the magnetometer inside the smartphone by using a magnet that was passed near the screen and by observing where the maximum value of the field was reached. We denoted that point as the origin O(0,0,0) cm. The arrangement of the Cartesian axes in the phone is as shown in Fig. 2. Then, we rotated the cell phone on the table until the value of the $B_x$ coordinate was zero. In this situation, the magnetic axis of the Earth was contained in the YZ plane of the phone. We drew some coordinate axes on a sheet of paper and then moved the magnet at 1-cm intervals with the symmetry axis aligned with the OX axis and then, we read the data, $B$, measured with the magnetometer. Fig. 1 shows the experimental setup for this part.

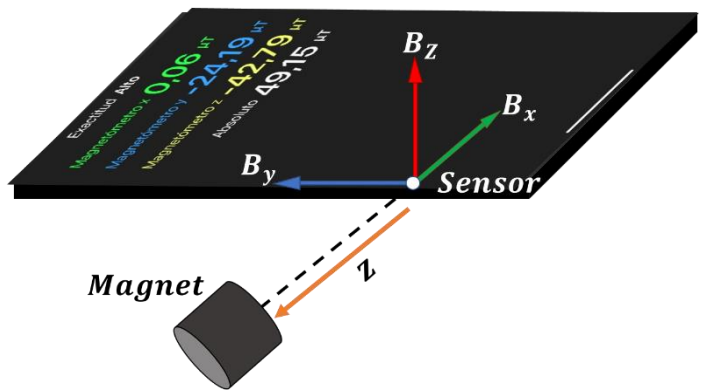

**Figure 1**. Experimental setup for the calibration with the coordinate system in the smartphone.

Eq. (1) can be written in a more compact form as $B(z) = kz^{-n}$, where $k$ is a positive parameter containing the characteristics of the magnet. Once found the value of $k$ we can isolate $m$ in the equation to determine its value:

$$m = \frac{k \cdot 2\pi}{\mu_0} \qquad (2)$$

Table 1 shows the collected data for the exponent calibration. For this purpose, the magnet was moved at intervals of 1 cm from $z_0 = 3$ cm (since for smaller distances the value of $B$ is larger than the top of the sensor range). The distances were taken with a house ruler of sensitivity $\Delta z = 0.1$ cm, and the sensor sensitivity that is reasonable to take is $\Delta B = 1$ µT, as the first decimal place fluctuates constantly. It should be added at this point that the sampling time of the sensor is one hundredth of a second ($\Delta t = 0.01$ s).

**Table 1.** Acquired data to calibrate the magnet.

| Distance | Field $B$ | Distance | Field $B$ |
|---|---|---|---|
| $z$ (cm) | $B$ (µT) | $z$ (cm) | $B$ (µT) |
| 3.0 | 2400 | 12.0 | 43 |
| 4.0 | 1200 | 13.0 | 34 |
| 5.0 | 575 | 14.0 | 27 |
| 6.0 | 338 | 15.0 | 22 |
| 7.0 | 207 | 16.0 | 18 |
| 8.0 | 142 | 17.0 | 16 |
| 9.0 | 100 | 18.0 | 13 |
| 10.0 | 72 | 19.0 | 10 |
| 11.0 | 55 | 20.0 | 10 |

With these experimental data, Fig. 2 is obtained. The fitting has been carried out using "LINEST" function of Excel. The use of Excel is convenient as MSOffice is commonly included in the syllabus of computing subjects in high schools, instead of more sophisticated software like Matlab, Mathematica or GNU Octave, more oriented to university students. The resulting value from the fitting for $n$ is $2.97 \pm 0.03$ which is compatible with the integer value $n = 3$ in Eq. 1. By performing uncertainty propagation in Eq. 2, we can obtain the uncertainty of the magnetic dipole moment which is finally expressed as $m = 0.330 \pm 0.015 \ A \cdot m^2$, consistent with values found in the literature (Arribas et al., 2015).

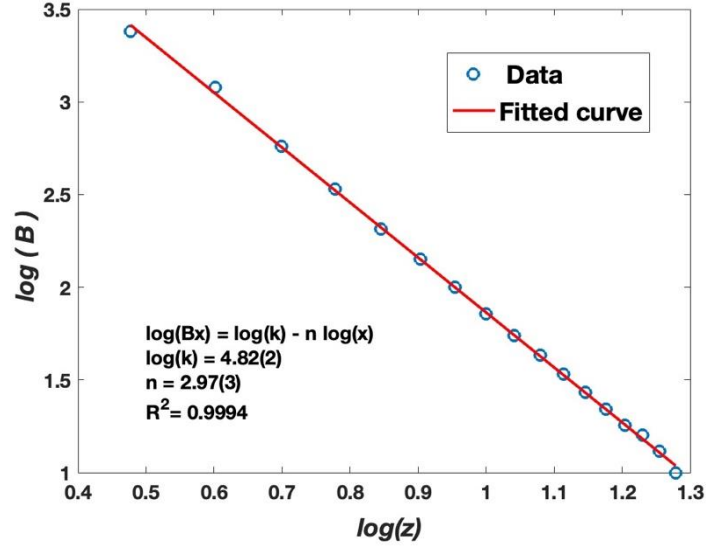

**Figure 2.** Fit of the measured values for *B* along its longitudinal symmetry axis

## 3. Study of an oscillating magnetic field along the axis for far and near conditions

Having already determined the value of $n$ in Eq. (2), we can leave it fixed and see how the field behaves at small distances from the sensor first ($z \sim L$) and then at considerably larger distances ($L \ll z$). For this purpose, we will attach the magnet to a spring of elastic constant $k = 20$ N/m and mass $m_{spring} = 2$ g (effective mass = 2/3 g); the total mass of the oscillating part is 208.7 g. Fig. 3 shows the experimental setup. For this part of the experiment, and to avoid external magnetic fields interfering with our oscillating magnet, other electric or electronic devices are kept away from the experimental setup. The equilibrium position $z_0$ will be set fixed, and then we will make the system oscillate, collecting the $B_z$ component of the field, which lies perpendicular to the smartphone screen. The exact expression expected for the magnetic field is given by the Eq. (3),

$$B = \frac{D}{(1+\xi)^3} \quad (3)$$

with $D = \mu_0 m / 2\pi z_0^3$ (see Eq. (1)), $\xi = \Delta z / z_0$, and $\Delta z = A \sin(\omega t + \varphi_0)$, being $A = \Delta z_{max}$ the oscillation amplitude, $\omega$ is the oscillation frequency, and $\varphi_0$ is the initial phase. To do the study, we will take data and fit them using the least squares method in Matlab. In this part of the experiment, in addition to $z_0$, the oscillation frequency is also assigned as obtained in $\omega = \sqrt{k/m} = 9.79$ rad/s. Both the fitted function and the different terms of the Taylor series given in Eq. (4) will be then represented. After expanding the function in Taylor series, Eq. (3) can be approached as follows:

$$B \approx D \sum_{j=0}^{S} \frac{1}{j!} \frac{d^j B}{d\xi^j}\bigg|_{\xi=0} \xi^j = D\left[1 - 3\xi + 6\xi^2 - 10\xi^3 + \ldots + \frac{1}{2}(-1)^S (S+1)(S+2)\xi^S\right] \quad (4)$$

where $S$ is the expansion order, so the number of terms in the Taylor series is $S+1$.

We demonstrate that only two terms in the Taylor series are necessary for the far field ($z_0 = 15$ cm), where the term $A/z_0 \ll 1$. However, for the near field ($z_0 = 3.5$ cm) more terms are necessary, as the condition above is no longer valid.

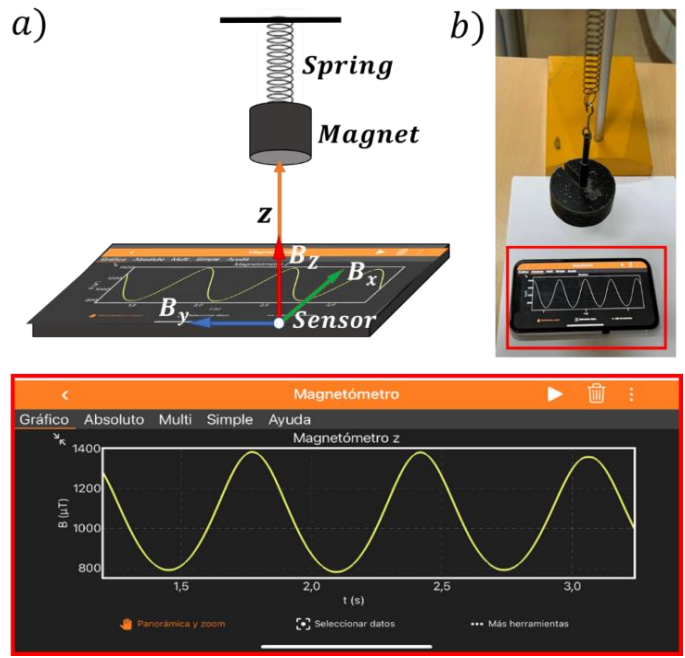

**Figure 3.** Experimental setup for the study of a long-axis magnetic field for far and near conditions, a) system string system and b) recording of oscillating $B$ magnetic field via smartphone.

For the study of the behaviour of the magnetic field $B$ in conditions far from the sensor, the magnet was placed at an equilibrium distance $z_0 = 15$ cm and made to oscillate. The fit to the theoretical data was implemented taking only the first two terms ($S = 1$), so $B \approx D(1 + \xi)$. The resulting plot can be seen in Fig. 4. The regression coefficient obtained for the data is $R^2 = 0.9971$, and the frequency found in the fit (this time, this parameter was set free) is $\omega = 9.76$ rad/s, in very good agreement with the theoretical value. The discrepancy between both values is 0.3%.

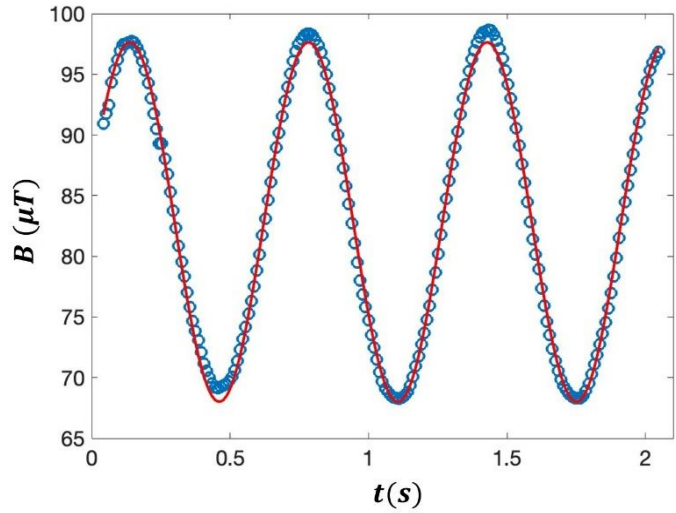

**Figure 4.** Fitting to experimental data (o) for the far field $B$ corresponding only to two terms in the Taylor series (red line).

On the other hand, under conditions of the magnet close to the sensor, the equilibrium distance has been set to $z_0 = 3.5$ cm, and it has been assumed that the exact function for the magnetic field is the one given in Eq. (3) where, in order to make the least squares fit, in MATLAB the values of $z_0$, $w$ and $n$ have been fixed, leaving the others free (offset, numerator $D$, amplitude $A$ and initial phase $\varphi_0$). In Fig. 5a), the original data are represented together with some other fittings. When approaching the second peak represented in the previous graph, it is possible to observe that the convergence of the series to the original function increases as the number of terms increases (see Fig. 5b).

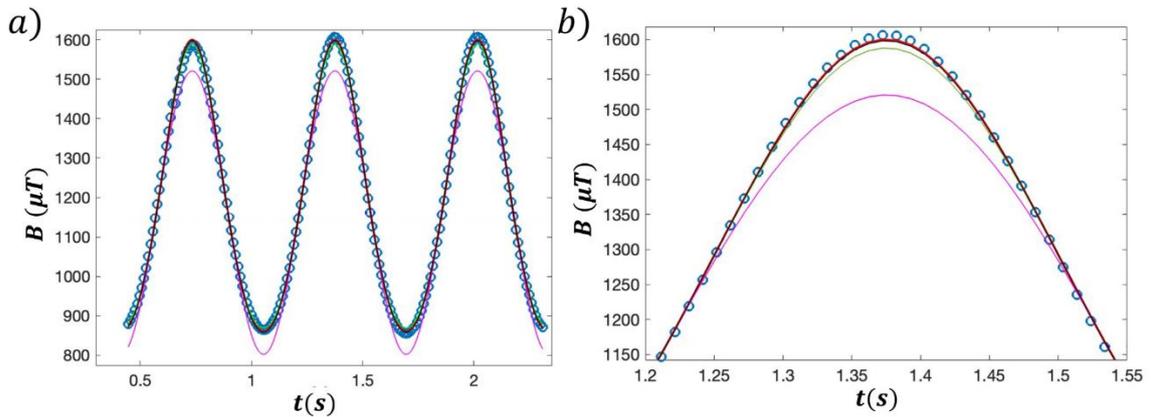

**Figure 5.** a) Fittings to experimental data (o) for the exact function (red line) and its Taylor expansions with $S = 1$ (pink line), $S = 2$ (green line), and $S = 3$ (grey line), respectively, and b) the approach to the second peak of the magnetic field plotted on the graph.

To measure the discrepancy between the recorded data and the estimation model, the sum of squared residuals (SSE) was calculated using Eq. (5),

$$SSE = \frac{1}{N}\sum_{i=1}^{N}(B_i - B_{S,i})^2 \qquad (5)$$

where $N$ the number of experimental data, $B_i$ the $i$-th experimental values and $B_{S,i}$ the corresponding values obtained with the approximated function, Eq. (4) of the expansion of order $S$. As shown in Table 2, as the number of terms increases, the SSE decreases, and the determination coefficient approaches to 1. In fact, with only four terms ($S=3$) the determination coefficient is $R^2 > 0.99$.

**Table 2.** Determination coefficient ($R^2$) and SSE for the different order $S$ in the Taylor series.

| Expansion order: $S$ | Determination coefficient: $R^2$ | SSE (($\mu T)^2$) |
|---|---|---|
| 1 | 0.956 | $2.08 \cdot 10^5$ |
| 2 | 0.9571 | $1.23 \cdot 10^2$ |
| 3 | 0.9909 | $5.36 \cdot 10^1$ |
| 4 | 0.9987 | $2.75 \cdot 10^{-2}$ |
| 5 | 0.9993 | $9.6 \cdot 10^{-3}$ |
| **Exact function** | 0.9993 | $5.32 \cdot 10^{-15}$ |

Furthermore, it can be concluded that as the expansion order ($S$) increases, the sum of squared errors (*SSE*) progressively decreases, approaching zero. This trend is clearly observed in Fig. 6, where the increase in terms in the series leads to higher model accuracy, reducing the discrepancies between observed and predicted values.

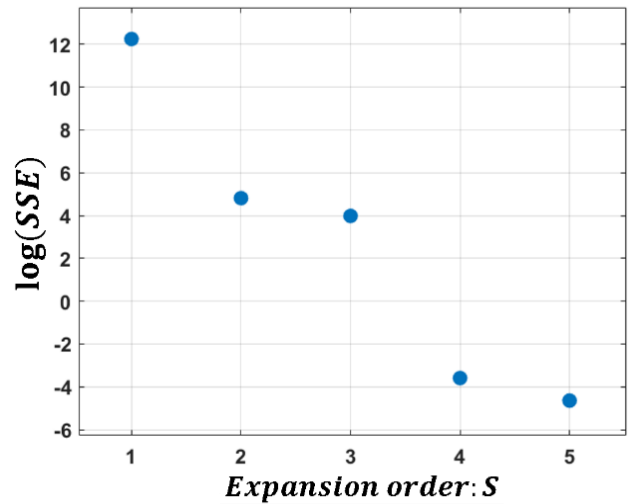

**Figure 6.** Representation of log(*SSE*) vs. number of terms ($1 \leq S \leq 5$), showing the error reduction as *S* increases.

## 4. Conclusions

After analyzing all data taken with the smartphone magnetometer, we can assume that this dependence of the field along the magnet axis is clearly shown as a cubic dependence for a cylindrical magnet. The modulus of the dipolar momentum of a small magnet can be estimated in an easy way by the students with a high level of accuracy and a relative error below 5%. The far-field B (under conditions of $A/z_0 \ll 1$) can be approached to a very good degree of agreement just by the first two terms in the Taylor series, however, the near-field B (where $A/z_0 \ll 1$ cannot be applied), at least four terms in the Taylor series must be taken to obtain a very good agreement with the best function matching the experimental data. Moreover, a very low-cost experiment (under 10 €) can be done by the students to calibrate the dependence of the field created by a magnet along its symmetry axis. Moreover, this experiment can be easily adapted for use at home, increasing the ease and motivation for students.

Mainly in Secondary Education, the modelling of physical phenomena using relevant mathematical equations is often undervalued in physics lectures which are typically heavily theoretical, focusing on problem-solving exercises with mostly closed and unique solutions.

In this context, it is necessary to raise the awareness of STEM teachers to introduce students to real Physics and Engineering problems that involve simulations of the physical phenomena involved, using the necessary mathematical equations that allow them to model the situation, changing different parameters and seeing how these affect the final solution of the problem.

This approach is essential for exposing students to open-ended, real-world scenarios, moving beyond the mere resolution of numerical exercises aimed solely at producing a final number and unit.

This article underscores the interdisciplinary nature of physics, engineering, mathematics, and ICT. It demonstrates how a familiar phenomenon like magnetism can be leveraged to design a simple, inexpensive experiment that illustrates the transition between two different situations (far and near fields). Both regimes can be effectively described using the powerful tool of Taylor series.

**Acknowledgements:** A.G.M. acknowledge the financial support from the Generalitat Valenciana (GRISOLIAP/2021/121). The authors thank the Instituto de Ciencias de la Educación de la Universitat Politècnica de València, Spain, for their support to the teaching innovation group MSEL.

**Author Cotributions:** All authors have contributed equally. All authors have read and agreed to the published version of the manuscript.

**Funding:** This work was supported by the Spanish Ministerio de Ciencia e Innovación (grant PID2022-142407NB-I00) and by Generalitat Valenciana (grant CIPROM/2022/30), Spain.

**Conflicts of Interest:** The authors declare no conflict of interest

**Data Availability / Supplementary Materials Statement:** All data generated or analyzed during this study are included in this article.

**References**

Arribas, E., Escobar, I., Suarez, C. P., Najera, A., & Beléndez, A. (2015). Measurement of the magnetic field of small magnets with a smartphone: A very economical laboratory practice for introductory physics courses. *Eur. J. Phys.*, *36*(6), 065002. https://doi.org/10.1088/0143-0807/36/6/065002.

Bissell, J. J. (2023). Proof of the small angle approximation sin $\theta \approx \theta$ using the geometry and motion of a simple pendulum. *Int. J. Math. Educ. Sci. Technol.*, 1–7. https://doi.org/10.1080/0020739X.2023.2258885.

Castro-Palacio, J.C., Velazquez-Abad, L., Gimenez, F. and Monsoriu, J.A. (2013a). Using a mobile phone acceleration sensor in physics experiments on free and damped harmonic oscillations, *Am. J. Phys. Educ., 81*, 472-475. https://doi.org/10.1119/1.4793438

Castro-Palacio, J.C., Velazquez-Abad, L., Gimenez, F. and Monsoriu, J.A. (2013b). A quantitative analysis of coupled oscillations using mobile accelerometer sensors. *Eur. J. Phys. Educ., 34,* 737–744. https://doi.org/10.1088/0143-0807/34/3/737

Dumrongkitpakorn P., Khemmani S., Plaipichit S., Wicharn S., & Puttharugsa C. (2023). Measuring the average velocity and acceleration of a moving object on an inclined plane using a magnetic sensor on a smartphone. *Phys. Educ.*, *58*, 013002. https://doi.org/10.1088/1361-6552/ac9e39.

Erol M., & Kara, A. (2023). Mathematical Modeling and Teaching of Outer Magnetic Field of a Solenoid Using Smartphones. *Phys. Educ., The*, 0*5(03)*, 2350012. https://doi.org/10.1142/S2661339523500129.

Giménez, M.H., Salinas, I. and Monsoriu, J.A. (2017). Direct Visualization of Mechanical Beats by means of an Oscillating Smartphone. *The Physics Teacher, 55,* 424-425. https://doi.org/10.1119/1.5003745

Gómez-Tejedor, J.A., Castro-Palacio, J.C. and Monsoriu, J.A. (2014). The acoustic Doppler effect applied to the study of linear motions. Introduction to linear motions. *European Journal of Physics Education, 35,* 025006. https://doi.org/10.1088/0143-0807/35/2/025006

Hochberg K., Kuhn J., & Müller A. (2018). Using smartphones as experimental tools—effects on interest, curiosity, and learning in physics education", *J. Sci. Educ. Technol.*, *27*, 385–403. https://doi.org/10.1007/s10956-018-9731-7.

Kuhn, J. and Vogt, P. (2013a). Analyzing acoustic phenomena with a smartphone microphone. *The Physics Teacher 51*, 118–119. https://doi.org/10.1119/1.4775539

Kuhn, J. and Vogt, P. (2013b). Applications and examples of experiments with mobile phones and smartphones in physics lessons. *Frontiers in Sensors 1*,4, 67-73.

Kuhn, S., and Vogt, M. (Eds.). (2018). Smartphones as mobile minilabs in physics. Springer.

Listiaji, P., Wulandari T.D., & Putri A.A. (2023). Application of the Oscillation Concept: Measuring the Human Respiration Rate in Various Activities Using a Smartphone's Magnetometer Sensor. *Phys. Teach.*, *61*, 304-306. https://doi.org/10.1119/5.0060098.

Marín-Sepúlveda, C., Ortuño-Molina, S., Castro-Palacio, J.C. and Monsoriu, J.A. (2024). Acoustic testing of the inverse-square law using the infrared signal of a remote control. *Physics Education, 59* (3), id.035016, 4 pp. https://doi.org/10.1088/1361-6552/ad37e7

Monteiro, M., Cabeza, C., & Martí, A. C. (2014). Angular velocity and centripetal acceleration relationship. *Phys. Teach., 52*(6), 389–391. https://doi.org/10.1119/1.4890502

Monteiro M., Stari C., Cabeza C., & Martí A.C. (2017). Magnetic field 'flyby' measurement using a smartphone's magnetometer and accelerometer simultaneously. *Phys. Teach.*, *55*, 580-581. https://doi.org/10.1119/1.5011840.

Ogawara Y., Bahri S., & Mahrley S. (2017). Observation of the magnetic field using a smartphone, *Phys. Teach.*, *55*, 184-185. https://doi.org/10.1119/1.4976667.

Parolin, S.O. and Pezzi, G. (2013). Smartphone-aided measurements of the speed of sound in different gaseous mixtures. *Phys. Teach., 51,* 508-509. https://doi.org/10.1119/1.4824957

Polak, R.D., Fudala, N., Rothchild, J.T., Weiss, S.E. and Zelek, M. (2016). Easily accessible experiments demonstrating interference. *Phys. Teach., 54,* 120-121. https://doi.org/10.1119/1.4940181

Riley, K. F., Hobson, M. P., & Bence, S. J. (2006). *Mathematical methods for physics and engineering* (3rd Edition). Cambridge University Press.

Salinas, I., Giménez, M.H., Monsoriu, J.A. and Castro- Palacio, J.C. (2018a). Characterization of linear light sources with the smartphone's ambient light sensor. *Phys. Teach., 56,* 562-563. https://doi.org/10.1119/1.5064575

Salinas, I. Giménez, M.H., Monsoriu, J.A. and Castro- Palacio, J.C. (2018b). El smartphone como barómetro en experimentos de Física. *Mod. Sci. Educ. Learn., 11*(1). https://doi.org/10.4995/msel.2018.9021

Sans J.A., Manjón, F.J., Pereira A.L.J., Gomez-Tejedor, J.A. and Monsoriu, J.A. (2013). Oscillations studied with the smartphone ambient light sensor. *Eur. J. Phys. 34* (6), 1349–1354. https://doi.org/10.1088/0143-0807/34/6/1349

Shakur, A. and Sinatra, T. (2013). Angular momentum. *Phys. Teach., 51,* 564-565. https://doi.org/10.1119/1.4830076

Staacks S., Hütz S., Heinke H., & Stampfer C. (2018). Advanced tools for smartphone-based experiments: phyphox. *Phys Educ., 53(4)*, 045009. https://doi.org/10.1088/1361-6552/aac05e.

Torriente-García, I., Muñoz, F.M., Castro-Palacio, J.C. and Monsoriu, J.A. (2023). RLC series circuit made simple and portable with smartphones. *Phys. Educ, 59,* 015016. https://doi.org/10.1088/1361-6552/ad04fb

Vieryas, (2020) Physics toolbox URL https://www.vieyrasoftware.net/ (accessed on 21-08- 2024).

Vogt, P., Kuhn J., (2022). Analyzing free fall with a smartphone acceleration sensor. *Phys. Teach., 50,* 182-183. https://doi.org/10.1088/1361-6552/ad04fb

Vogt, P., Kuhn J., (2022). Analyzing simple pendulum phenomena with a smartphone acceleration sensor. *Phys. Teach., 50*, 439- 440. https://doi.org/10.1119/1.4752056

Wannous, J. & Horvath, P. (2023). Precise measurements using a smartphone's magnetometer. Measuring magnetic fields and permeability. *Phys. Teach.*, *61*, 36-39. https://doi.org/10.1119/5.0033597.

Westermann, N., Staacs S., Heinke H., & Möhrke P. (2022). Measuring the magnetic field of a low frequency LC-circuit with phyphox. *Phys. Educ.*, *57(6)*, 065024. https://doi.org/10.1088/1361-6552/ac920e.